\documentclass[aps,superscriptaddress,amssymb,twocolumn]{revtex4}
\usepackage{float}
\usepackage{graphicx}
\usepackage{epstopdf}
\usepackage[T1]{fontenc}
\usepackage[latin9]{inputenc}
\usepackage{amsbsy}
\usepackage{amsmath}
\usepackage{color}
\DeclareMathOperator{\sech}{sech}

\newcommand{\pnr}[1]{}
\newcommand{\pnb}[1]{}
\newcommand{\png}[1]{}

\begin{document}


\title{
Anomalous thermodynamic power laws near topological transitions in nodal superconductors
}



\author{B. Mazidian}
\affiliation{H. H. Wills Physics Laboratory, University of Bristol, Tyndall Avenue, Bristol BS8 1TL, UK}
\affiliation{ISIS facility, STFC Rutherford Appleton Laboratory, Harwell Science and Innovation Campus, Oxfordshire, OX11 0QX, UK}
\author{J. Quintanilla}
\email[]{j.quintanilla@kent.ac.uk}
\affiliation{SEPnet and Hubbard Theory Consortium, School of Physical Sciences, University of Kent, Canterbury CT2 7NH, UK}
\affiliation{ISIS facility, STFC Rutherford Appleton Laboratory, Harwell Science and Innovation Campus, Oxfordshire, OX11 0QX, UK}
\author{A.D. Hillier}
\affiliation{ISIS facility, STFC Rutherford Appleton Laboratory, Harwell Science and Innovation Campus, Oxfordshire, OX11 0QX, UK}
\author{J. F. Annett}
\affiliation{H. H. Wills Physics Laboratory, University of Bristol, Tyndall Avenue, Bristol BS8 1TL, UK}


\date{\today}

\begin{abstract}
Unconventional superconductors are most frequently identified
by the observation of power-law behaviour on low-temperature thermodynamic
or transport properties, such as specific heat. 
Here we show that, in addition to the usual point and line nodes, a much wider class of different nodal types
can occur. These new types of nodes typically occur when there are transitions between different types of gap node topology, for example when point or line nodes first appear as a function of some physical parameter. We identify anomalous, non-integer thermodynamic power laws associated with these new nodal types, and give 
physical examples of 
superconductors in which they might be observed experimentally, including the noncentrosymmetric superconductor Li$_2$Pd$_{3-x}$Pt$_x$B.
\end{abstract}



\maketitle


A defining feature of many unconventional Fermi superfluids and superconductors is the existence of lines or points on the Fermi surface where the energy gap vanishes and so-called "nodal quasi-particles" can exist at arbitrarily low energies. 
In Fermi systems with such nodal quasi-particles
the low temperature specific heat will show particular power law behaviours as a function of temperature. 
The expected point and line node power law dependencies were first derived in relation to the proposed low temperature superfluid phases of liquid $^{3}$He  \cite{And_Morel_61}. These were subsequently clarified further  \cite{Leggett79} 
and are now widely used to identify pairing states in unconventional superconductors \cite{SigristUeda89,Annett90,MacKenzieMaeno2003}.
Here we show that gap nodes in superconductors can occur in a number of more general types than simply the usual line or point zeros and that each of these has a corresponding thermodynamic signature, 
typically 
in the form of non-integer power laws in low temperature
specific heat.  We predict that these anomalous power laws generically occur at points in the phase diagram where there is a topological change in the line or point nodal structure
 on the Fermi surface, which we illustrate with a specific example: the noncentrosymmetric superconductors Li$_2$Pd$_{3-x}$Pt$_x$B.   
In this case the gap node topological changes are also associated with changes of
bulk topological quantum numbers for the quasi-particles. 
The future experimental observation of such non-integer power laws could 
therefore be used to identify not only superconductors with highly unconventional pairing symmetries,
but also topological superconducting and superfluid systems \cite{topologicalsuperconductors}. Similar physics are realised in the high-temperature cobalt doped Pnictide materials Ba(Fe$_{1-x}$T$_x$)$_2$As$_2$ (T=Co,Ni,Pd) \cite{Fernandes2011, Stanev2011, Mishra}.



\begin{figure}
\begin{tabular}{|c|c|}
\hline 
\textbf{a) linear point} & \textbf{b) shallow point}  \tabularnewline
\textbf{node} & \textbf{ node } \tabularnewline
\includegraphics[trim = 14mm 14mm 14mm 9mm, clip, width=0.3\columnwidth]{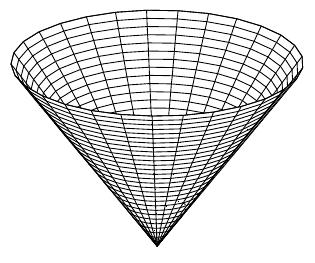}  & 
\includegraphics[trim = 14mm 14mm 14mm 9mm, clip, width=0.3\columnwidth]{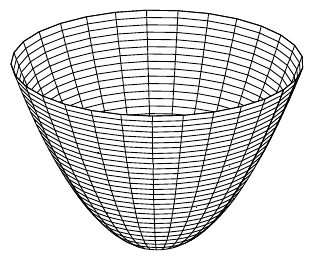}\tabularnewline
$|\Delta_{k}|^{2}=I_{1}\left({k_{x\parallel}}^{2}+{k_{y\parallel}}^{2}\right)$ &
 $|\Delta_{k}|^{2}=I_{1}\left({k_{x\parallel}}^{2}+{k_{y\parallel}}^{2}\right)^{2}$\tabularnewline
$\begin{array}{rcl}
g\left(E\right) & = & \frac{E^{2} }{2(2\pi)^{3}I_{1}\sqrt{I_{2}}}
\end{array}$ & $\begin{array}{rcl}
g(E) & = & \frac{E}{2(2\pi)^{3}\sqrt{I_{1}I_{2}}}
\end{array}$\tabularnewline
$n=3$ & $n=2$ \tabularnewline
\hline 
\textbf{c) linear line} & \textbf{d) shallow line}  \tabularnewline
\textbf{node} & \textbf{node}  \tabularnewline
\includegraphics[trim = 9mm 10mm 10mm 8mm, clip, width=0.3\columnwidth]{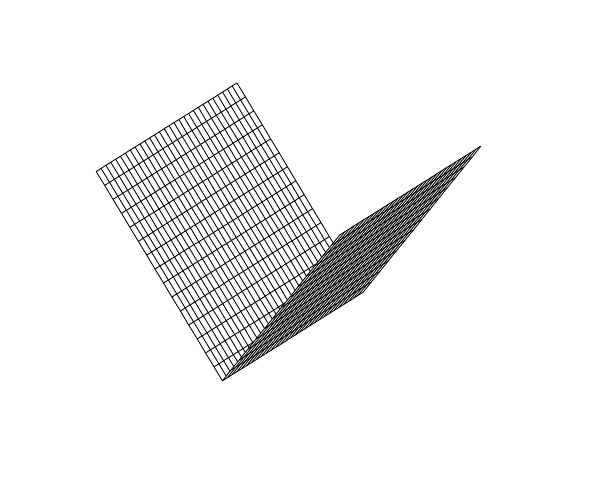}  &
\includegraphics[trim = 9mm 10mm 10mm 8mm, clip, width=0.3\columnwidth]{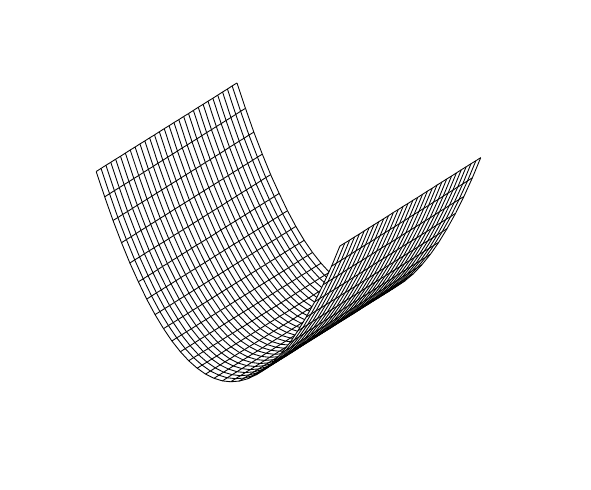}\tabularnewline
$|\Delta_{k}|^{2}=I_{1}{k_{x\parallel}}^{2}$ & $|\Delta_{k}|^{2}=I_{1}{k_{x\parallel}}^{4}$\tabularnewline
$\begin{array}{rcl}
g\left(E\right) & = & \frac{LE}{(2\pi)^{3}\sqrt{I_{1}I_{2}}}
\end{array}$ & $\begin{array}{rcl}
g(E) & = & \frac{L\sqrt{E}}{(2\pi)^{3}I_{1}^{\frac{1}{4}}\sqrt{I_{2}}}
\end{array}$\tabularnewline
$n=2$ & $n=1.5$ \tabularnewline
\hline 
\textbf{e) crossing} & \textbf{f) crossing}\tabularnewline
\textbf{ of linear line nodes } & \textbf{of shallow line nodes}\tabularnewline
\includegraphics[trim = 9mm 8mm 10mm 8mm, clip, width=0.3\columnwidth]{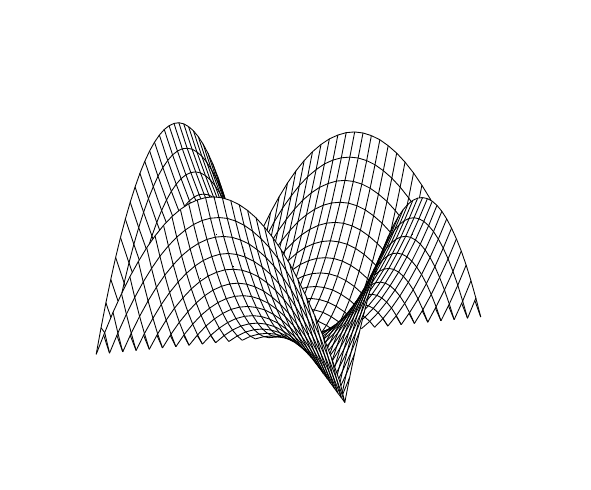} & 
\includegraphics[trim = 9mm 8mm 10mm 8mm, clip, width=0.3\columnwidth]{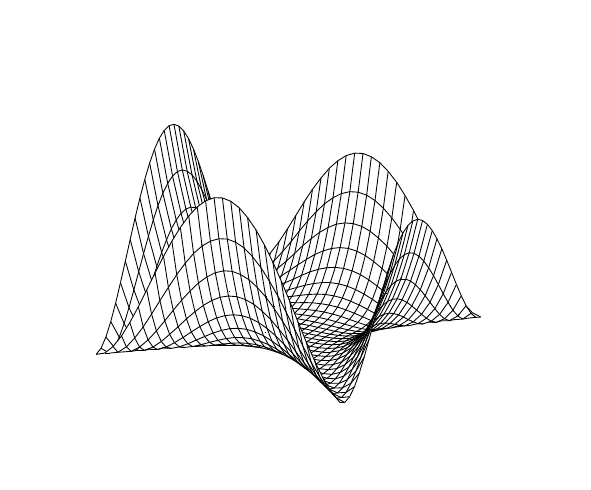}
\tabularnewline
 $|\Delta_{k}|^{2}=I_{1}\left({k_{x\parallel}}^{2}-{k_{y\parallel}}^{2}\right)^{2}$ & 
 $|\Delta_{k}|^{2}=I_{1}\left({k_{x\parallel}}^{2}-{k_{y\parallel}}^{2}\right)^{4}$\tabularnewline
or $I_{1}{k_{x\parallel}}^{2}{k_{y\parallel}}^{2}$ & or 
$I_{1}{k_{x\parallel}}^{4}{k_{y\parallel}}^{4}$\tabularnewline
 $\begin{array}{rcl}
g(E) & = & \frac{E(1+2\ln|\frac{L+\sqrt{E}/I_{1}^{\frac{1}{4}}}{\sqrt{E}/I_{1}^{\frac{1}{4}}}|)}{(2\pi)^{3}\sqrt{I_{1}I_{2}}}\\
 & \sim & E^{0.8}
\end{array}$& $\begin{array}{rcl}
g(E) & = & \frac{\sqrt{E}(1+2\ln|\frac{L+E^{\frac{1}{4}}/I_{1}^{\frac{1}{8}}}{E^{\frac{1}{4}}/I_{1}^{\frac{1}{8}}}|)}{(2\pi)^{3}I_{1}^{\frac{1}{4}}\sqrt{I_{2}}}\\
 & \sim & E^{0.4}
\end{array}$\tabularnewline
 $n=1.8$ & $n=1.4$\tabularnewline
\hline 
\end{tabular}
\caption{\label{Fig:points_lines_nodal_crossings} The four generic types of point and line nodes 
in a superconductor or Fermi superfluid and the two generic types of line crossings: (a) a point node with a linear excitation spectrum,
(b) a shallow point node, (c) a line node with linear spectrum, (d) a shallow line node, (e) a crossing of ordinary line nodes, and
(f) a crossing of shallow line nodes.  
For the point and line nodes (a-d) we give example gap functions,  densities of states and low-$T$ specific heat capacity
power law exponents  $n$ for $C_{v} \propto T^n$.  The coordinates $k_{x\parallel}$ and $k_{y\parallel}$
refer to a local coordinate system on the Fermi surface centered on the node, where $k_z$ is normal to the Fermi surface,
and where $k_{x\parallel}$ and $k_{y\parallel}$ are coordinates tangential to the plane of the Fermi 
surface at the node. The crossings (e,f) each have a distinct gap spectrum, 
density of states and power law which is distinct from the cases of point and line nodes alone,
and which becomes dominant at low temperatures. 
The densities of states in these two cases are logarithmic, where the parameter $L$ denotes
 the length of the line nodes on the Fermi surface (measured to midway between crossing points). 
 These densities of states approximate a power law, as shown, and we also show the approximate heat capacity
power law exponents $n$. Details of the derivations are given in Appendix~\ref{sec:methods}.}
\end{figure}

For the usual types of point and line nodes the quasi-particle energy spectrum is linear
in the vicinity of the gap node. A familiar example is the case of the $^{3}$He A-phase 
with triplet pairing order parameter ${\bf d}_{\bf k}\propto (k_x+ik_y)\hat{\bf z}$, which has point nodes
at the points ${\bf k}=(0,0,\pm k_F)$ on the Fermi sphere, of radius $k_F$. Near to these
points the quasi-particle energy spectrum, $E_{\bf k}$, obeys $E_{\bf k}^2=v_F^2 (k_z-k_F)^2 + |{\bf d}_{\bf k}|^2$, 
where $v_F$ is the Fermi velocity, giving a linear
dispersion relation as shown in Fig. \ref{Fig:points_lines_nodal_crossings} a). 
For this type of gap node
 we expect a specific heat, $C_{v} \propto T^{3}$ at temperatures $T$ much lower than the 
critical temperature $T_c$.  The  case of line nodes in the superconducting gap is also well known, as for example found in d-wave superconductors such as YBa$_2$Cu$_3$O$_7$. Assuming in this case that
 the Fermi surface is an open cylinder of radius $k_F$, and the energy gap is of the form 
 $\Delta_{\bf k}= \Delta_0 (k_x^2-k_y^2)$, then there are four line nodes spaced around the Fermi 
 surface at ${\bf k}=
k_F (\pm 1,\pm 1)/\sqrt{2}$. Again the energy spectrum is linear near these gap nodes, as shown in 
Fig. \ref{Fig:points_lines_nodal_crossings} c), so in this case we expect $C_{v} \propto T^{2}$ \cite{Hussey2002}.


While the generic point and line nodes of this type are expected in the majority of unconventional 
superconductors,  other types of nodal energy spectrum are possible. For example it is possible that the system has a point node, but that the energy spectrum is quadratic not linear near 
the nodal point --see Fig. \ref{Fig:points_lines_nodal_crossings} b). The heat capacity at this critical point
is a power law, with $C_v \propto T^2$ rather than the usual $C_{v} \propto T^{3}$ for point nodes \cite{Leggett79}.  This aspect of non-linear excitation spectra was identified by Leggett in his review of the superfluid phases of $^3$He \cite{Leggett79}, though at the time no examples were known. Later a ground state with E$_{2u}$ symmetry  having this feature was proposed for the heavy-fermion superconductor UPt$_3$  \cite{Sauls94} and possible thermal transport signatures were discussed \cite{UPt3_thermal}. Here we show that this different type of point node behaviour is not restricted to certain special symmetries but 
occurs generically when the gap structure is changing from having a point node to having no nodes, as a function of 
some parameter defining the pairing potential on the Fermi surface. 
We extend this notion to a much wider class of nodal types and node topology transitions. 

The fact that it is possible to continuously remove a point node of the gap is surprising, since
the usual linear spectrum point node behaviour is usually thought to be topologically stable against
perturbations \cite{Volovik,Volovik2,Sato,Volovik3}.
The shallow node behaviour is not
topologically protected, since it only occurs for specific parameter values in which 
accidental cancellations occur between different order parameter components. Nevertheless, as we shall see its influence extends throughout the finite-temperature phase diagram.

If we generalize the shallow point node behaviour to line nodes we obtain a spectrum near to the node shown in
Fig. \ref{Fig:points_lines_nodal_crossings} d), which we will refer to as a \emph{shallow line node}. In this case we find
an anomalous non-integer power law for the low temperature specific heat, of the form 
 $C_{v}$ $\propto$ $T^{1.5}$.  Similarly to the shallow point node, this type of shallow line 
 node only occurs at a boundary between 
 a nodeless superconductor and one with  ordinary line nodes, as discussed recenly for the pnictide superconductors \cite{Fernandes2011, Stanev2011}.   In particular, both nodal and nodeless compounds of the system Ba(Fe$_{1-x}$Co$_{x}$)$_{2}$As$_{2}$  have been reported \cite{Gofryk10, Tanatar10}, suggesting that the $T^{1.5}$ power law could be observed experimentally. This would provide compelling evidence for the $s^{\pm}$ gap model \cite{Hirschfeld_2011}.

Another type of nodal behaviour occurs when there is a crossing or topological reconnection of line nodes. 
These again lead to distinctive low temperature power laws, with non-integer power law exponents. 
A crossing of linear line nodes is illustrated in Fig.~\ref{Fig:points_lines_nodal_crossings} (e). This type of
crossing can occur in a d-wave superconductor, for example $\Delta_{\bf k} \propto k_x^2-k_y^2$, in the case
where the Fermi surface is spherical rather than cylindrical. Although the node crossing point 
occurs for a small portion of the Fermi surface, it turns out that this point dominates the low temperature
heat capacity, and  such crossings can therefore be detected experimentally. This is because the
density of states resulting from the node crossing has a logarithmic term. This logarithmic
term approximates a power law at low temperatures, and leads to a specific heat capacity 
contribution approximately of the form $C_v \propto T^{1.8}$, which at low enough temperatures 
dominates over the usual line node
contribution $C_v \propto T^{2}$ arising from the line nodes away from the crossing. Unlike the shallow point node, the crossing of line nodes is expected to be topologically protected against small perturbations.

The final distinct type of gap node which we shall discuss is a crossing of shallow line nodes,
as shown in Fig. \ref{Fig:points_lines_nodal_crossings} (f). Again this leads to a logarithmic term in the
density of states, which dominates the low temperature behaviour compared to the shallow 
line node case in the absence of crossings.
 The corresponding specific heat capacity is approximately $C_v \propto T^{1.4}$ for the shallow line node
 crossing, compared to $C_v \propto T^{1.5}$ for the case without crossings. Interestingly, $n=1.4$ is not so far from the value obtained for the ungapped Fermi surface of a normal (non-superconducting) Fermi liquid, $n=1$.

\begin{figure}
\includegraphics[trim = 6mm 0mm 35mm 3mm, clip, width=0.7\columnwidth]{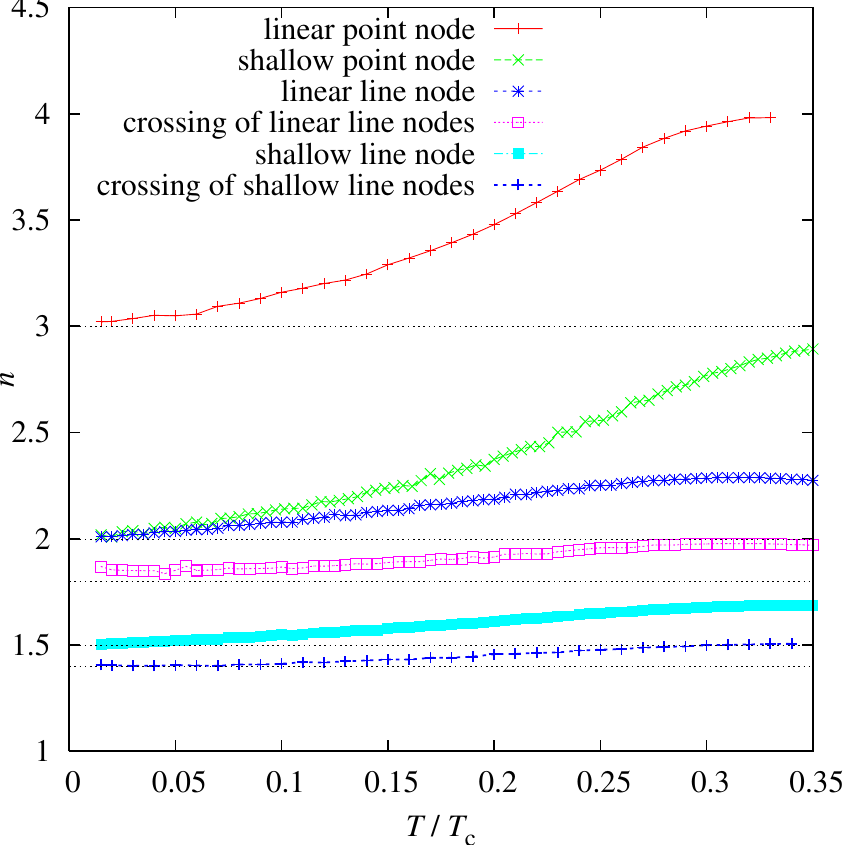}
%
\caption{\label{Fig:exponent_plot}(color online) The simulated exponent of temperature $T$ in the specific heat capacity as a function of temperature for each of the nodal characters. The value of $n$ in the power law $C_v \propto T^{n}$ is obtained by calculating $n=\frac{d\ln(C_v)}{d\ln(T)}$. These match analytics (dotted lines) in the limit $T \rightarrow 0$. The methodology employed to produce this plot is described in Appendix \ref{sec:methods}.}
\end{figure}

 Fig. \ref{Fig:exponent_plot} shows the predicted exponents of the power law temperature dependences
 of specific heat for each of the different nodal types shown in Fig.~\ref{Fig:points_lines_nodal_crossings}. In all cases the predicted power law
 behaviour $C_v \propto T^n$ is accurate at very low temperatures, below $0.1 T_c$,
 and in many cases the low temperature power laws apply over a range of temperatures
 up to $0.3 T_c$. The details of our calculations of specific heat leading onto the results in Figs.~\ref{Fig:points_lines_nodal_crossings} and \ref{Fig:exponent_plot} are given in Appendix \ref{sec:methods}.

 Non-integer power laws, especially powers below $2$, are a highly distinctive
 signature of the low energy excitation spectrum, and their experimental observation 
 would be a relatively clear and direct form of evidence for the existence of 
 the corresponding type of gap node in a given material.  In systems where the 
 gap topology is changing as a function of some experimental parameter (e.g.~doping
 or spin-orbit coupling) these anomalous powers will only be observed to 
 the lowest temperatures if the material is exactly at the critical parameter value for the 
 shallow node to occur. However, if the material is merely near to, not exactly at, the critical parameter
 value, then the shallow node power laws will hold over a wide temperature range, reverting to 
 ordinary line, point, or nodeless behaviour at the very lowest temperatures. Again experimental
 observation of such a cross-over between power laws would be a clear signature of the 
 unusual gap nodal structure - we provide an explicit example of this below.
 In practice, it is unlikely that the power laws shown in Fig. \ref{Fig:exponent_plot} can be distinguished from one-another in a single measurement as a function of temperature, however a transition of the type shown in Fig.~\ref{Fig:topology_plot_c} (inset), discussed below, is more feasible as a function of doping at constant temperature.

\begin{figure}
\includegraphics[width=1\columnwidth]{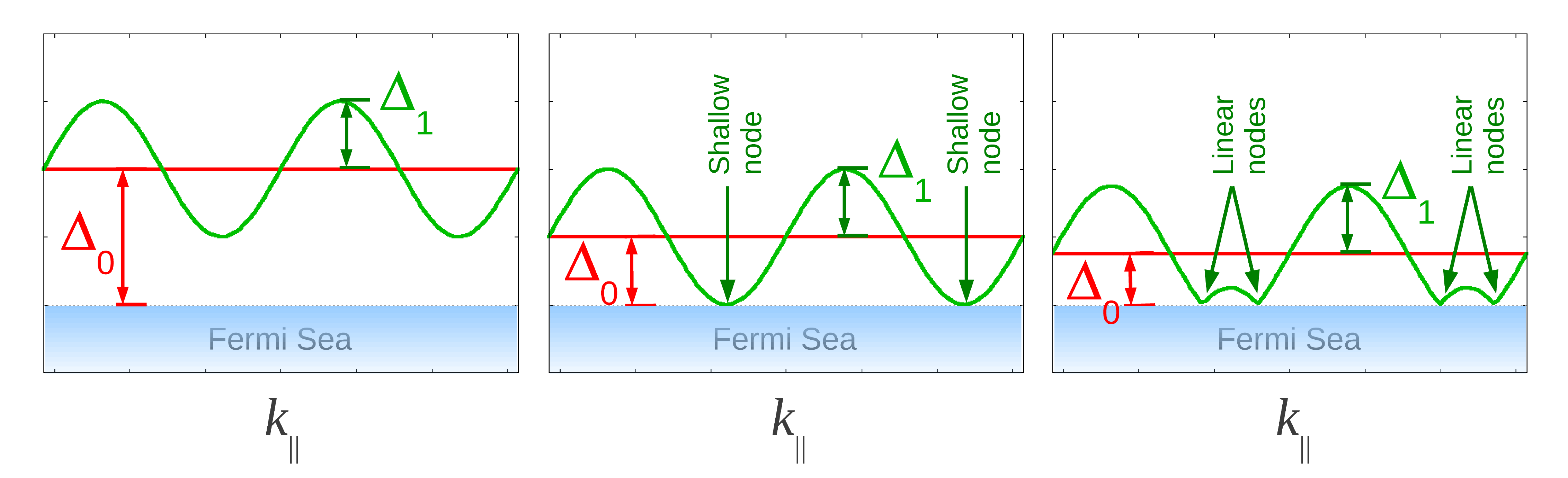}
\caption{\label{Fig:multi_param}(color online) The transition between a nodal and non-nodal gap structure in any superconductor as a result of continuously admixed order parameters will be characterized by a non-linear gap energy spectrum close to the node. This is because in any wave-like order parameter the deep minima have no linear order term in the expansion.}
\end{figure}

In addition to the {\it ad hoc} mechanisms mentioned above, we expect shallow line or point nodes to occur generically in superconductors with multi-component order parameters. 
Consider the simple model
shown in Fig. \ref{Fig:multi_param}. This shows a hypothetical
system, such as an ordinary s-wave BCS superconductor with a significant level of
anisotropy
 in the gap. Here $\Delta_0$ is the average gap and $\Delta_1$ is a measure of
the gap variation with $k_\parallel$ parallel to the Fermi surface. In the case
where $|\Delta_0| > |\Delta_1|$ the gap is nodeless everywhere on the Fermi surface.
On the other hand, if $|\Delta_0| < |\Delta_1|$  then the gap changes sign at some points on
 the Fermi
surface, leading to pairs of line nodes. At the critical parameter value
$|\Delta_0| = |\Delta_1|$ the line node pairs coalesce, leading to a single, shallow, line node.
In this example no symmetry change occurs in the gap function, however ``accidental'' line nodes
occur within a single symmetry of the Cooper pair order parameter, because of the large
degree of gap anisotropy. 

\begin{figure}
\includegraphics[angle=0,width=1\columnwidth]{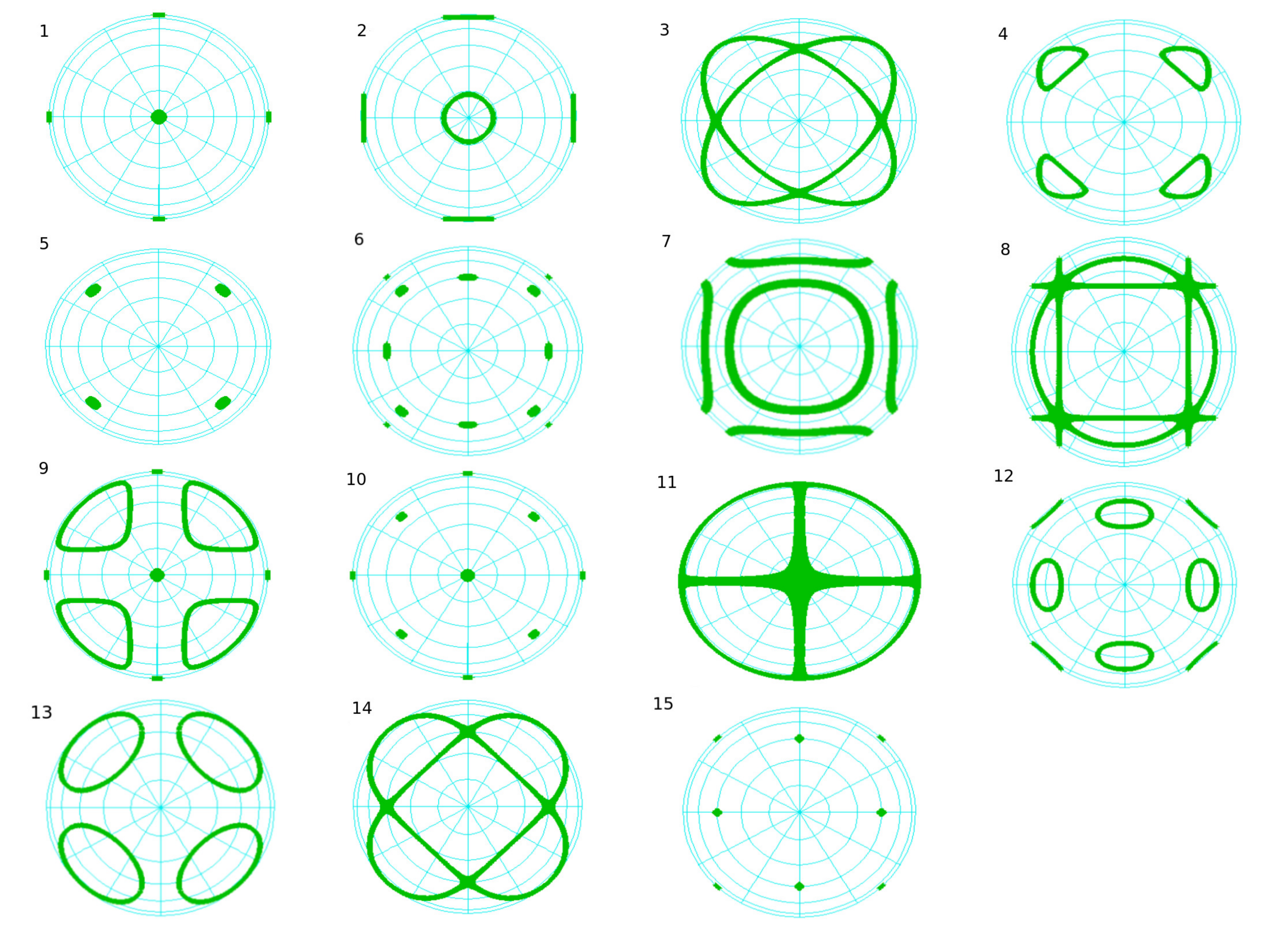} \\
\includegraphics[angle=270,width=1\columnwidth]{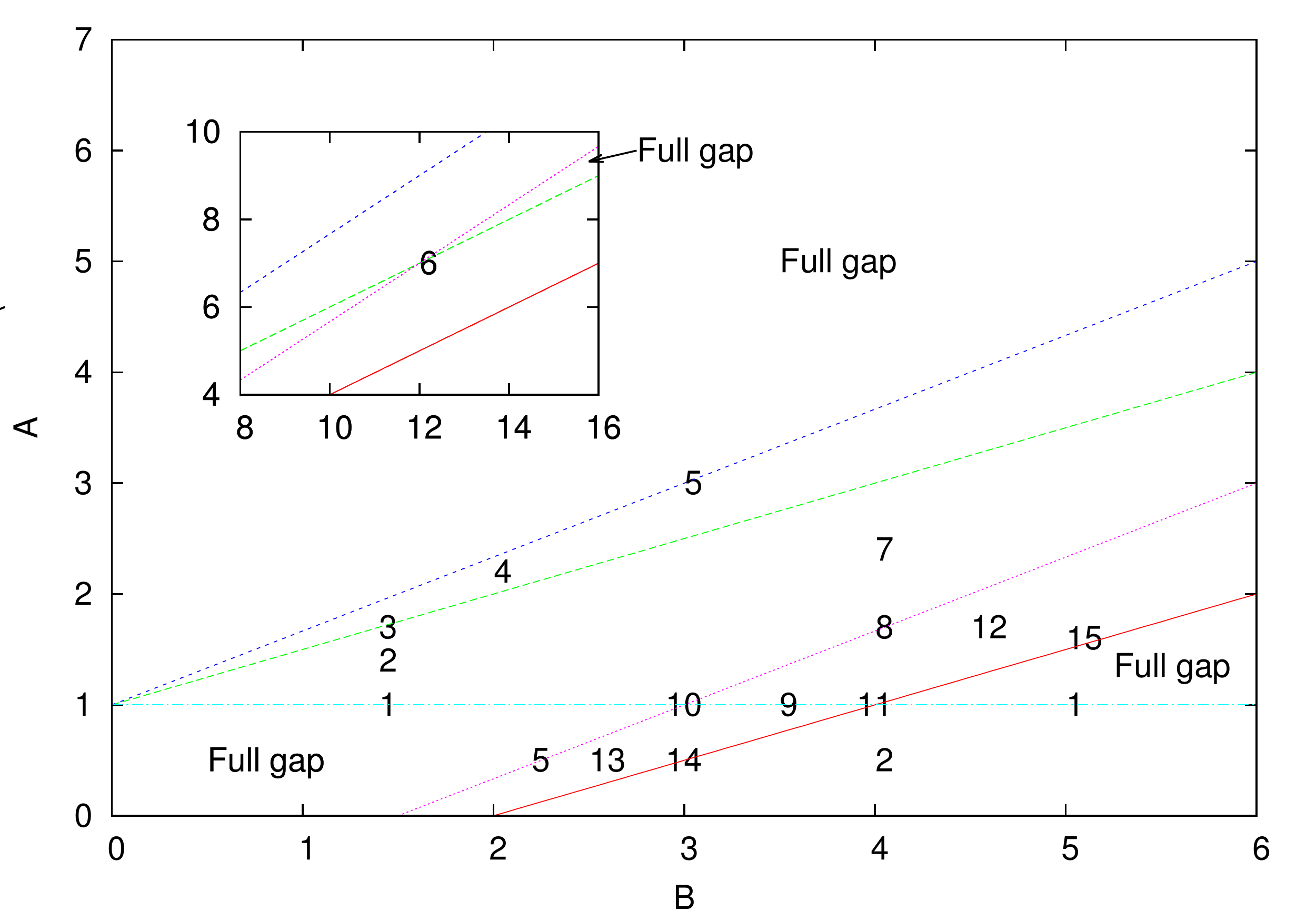} 
\caption{\label{Fig:topology_plot_a_b}(color online) (top) Fifteen gap topologies corresponding to different 
 triplet admixtures in a cubic non-centrsymmetric superconductor. Within
 this set the only point nodes are shallow point nodes, occuring in plots 1,5 10 and 15. 
 Topologies 3,6,8 and 14 are all line reconnection transitions between distinct topologies 
 and include crossings of linear line nodes. Topology 11 is a crossing of
  shallow line nodes. (bottom) Phase diagram of gap node topology corresponding to triplet admixture parameters $A$ and $B$ in Eq.~(\ref{LiPdPtB-gap-2}) on a spherical Fermi surface. 
Each line represents a phase boundary where gap nodes appear or disappear on the Fermi surface.
}
\end{figure}
\begin{figure}
\includegraphics[angle=0,width=1\columnwidth]{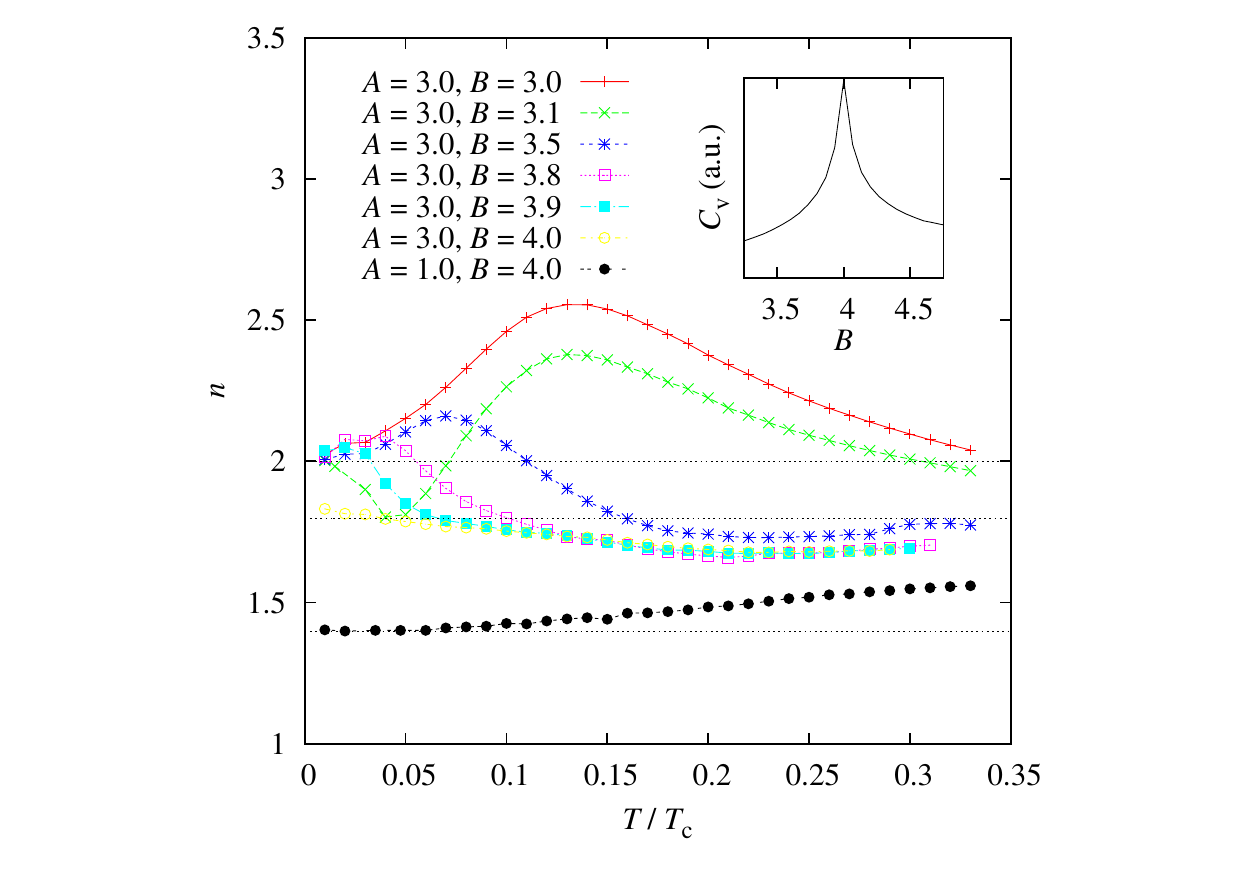}
\includegraphics[angle=0,width=1\columnwidth]{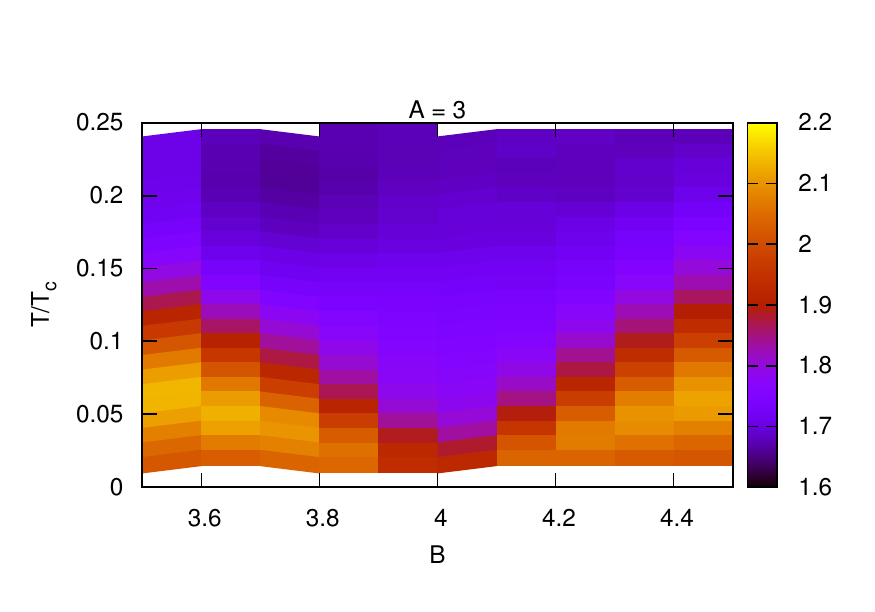}
\caption{\label{Fig:topology_plot_c}(color online) 
(top panel) Specific heat exponent as a function of temperature for different values of the $A$ and $B$ parameters, as indicated. The inset shows the value of the specific heat as a function of $B$ for fixed $A=3$ and constant temperature, $T=0.005~T_c$. (bottom panel) Specific heat exponent as a function of temperature and parameter $B$ for fixed $A=3$. Details of the calculation of the quasiparticle spectrum and phase diagram are given in Appendix \ref{sec:methods}.}
\end{figure}

 An explicit example of this  mechanism by which anomalous power laws may be observed 
 is provided by non-centrosymmetric superconductors where the superconductor order parameter 
\begin{equation}
 \hat{\Delta}_{\bf k}=\left(\Delta_{0,{\bf k}}+{\bf d}_{\bf k}\cdot \boldsymbol{\sigma}\right) i\hat{\sigma}_y
\end{equation} 
 has singlet and triplet components $\Delta_{0,{\bf k}}$ and ${\bf d}_{\bf k},$
  respectively, 
such as  
  the compounds Li$_{2}$Pd$_{3-x}$Pt$_{x}$B~\cite{Tog_Badi_04,Badi_Kon_05,Yuan2006}. Platinum doping in this system is thought to 
increase the spin-orbit coupling, and hence increase the
triplet component of the order parameter relative to the singlet component~\cite{Yuan2006,Nishiyama2007,Takeya2007,Harada2012}. 
In doing so the spin-orbit coupling introduces nodes in the gap. Within a standard
symmetry classification of the possible order parameter structures the simplest one takes the form \cite{Schnyder2011,Schnyder2012}
\begin{eqnarray}
\Delta_{0,\bf k}&=&\Delta_0 \label{LiPdPtB-gap-1} \\
 {\bf d}_{\bf k}&=&\Delta_0 \left\{ A  \left[ X,Y,Z \right]  - \right. \label{LiPdPtB-gap-2} \\
 && B \left. \left[ X\left(Y^2+Z^2\right) , Y\left(X^2+Z^2\right) , Z\left(Y^2+X^2\right) \right] \right\} \nonumber 
\end{eqnarray}
where $X,Y,Z$ are functions with the same symmetry properties as the components of the electron wave vector $k_x,k_y,k_z$ and $A$ and $B$ are material-specific parameters that depend on doping $x$. Eqs. (\ref{LiPdPtB-gap-1}) and (\ref{LiPdPtB-gap-2}) correspond to the $A_1$ irreducible representation of the cubic $O$ group, which is the only one where only gauge symmetry is broken. Other pairing states break additional symmetries and therefore are not compatible with a conventional (phonon-mediated) pairing mechanism. In spite of this, a large number of 
different gap nodal structures are obtained, as shown in Fig.~\ref{Fig:topology_plot_a_b}~(top panel).  As a function of the parameters $A,B$ we obtain the rich phase diagram of Fig.~\ref{Fig:topology_plot_a_b}~(bottom panel). The corresponding specific heat exponents are illustrated in Fig.~\ref{Fig:topology_plot_c}. Upon doping, the material explores the phase diagram along some unspecified path from small values of $A$ and $B$ ($x=0$) to larger values ($x=3$) \cite{Yuan2006}.
Small values of $A$ and $B$ correspond to a fully gapped, singlet-dominated superconducting order parameter as believed to be realised in Li$_{2}$Pd$_{3}$B. Order parameters with a large triplet component and line nodes on the Fermi surface are thought to occur in Li$_{2}$Pt$_{3}$B and correspond to the case when either $A$ or $B$ (or both) are larger.

Inspection of the phase diagram shows that at some
 values of $A$ and $B$ we 
expect point nodes to appear,  labelled by `1' in Fig.~\ref{Fig:topology_plot_a_b}, 
 before line nodes appear on the gap, labelled by `2' in the same figure. 
These point nodes are shallow point nodes, and so 
instead of the usual 
$T^{3}$ power law dependence of the heat capacity for point nodes, we instead expect a $T^{2}$ 
power law dependence. 
 Upon further change of the $A$ and $B$ parameters 
these shallow point nodes expand into rings of ordinary line nodes, as shown in Fig.~\ref{Fig:topology_plot_a_b}, plot labelled `2'. Increasing the parameters yet again there is a level which causes a topological line 
reconnection, labelled `3', where  the different line nodes
 intersect before becoming rings again, now centred at different points on the 
Fermi surface, `4'. Note that the number of line nodes on the Fermi surface is always a multiple of two as predicted by the topological theory of gapless phases in time-reversal-invariant superconductors \cite{Beri2010}. That number jumps at the line reconnection transition, indicating its topologically non-trivial nature. The line reconnection transition has a 
 distinct thermal signature 
in the low $T$ heat capacity originating from the crossing of line nodes associated 
with such a line reconnection. As is evident from the figure, there are many other paths along the phase diagram with similar phase transitions.
In our analysis of the cubic non-centrosymmetric superconductors we also found a single instance 
of crossing of shallow line nodes, labelled `11' in Fig.~\ref{Fig:topology_plot_a_b}. These are 
realised when line nodes are spontaneously introduced on a nodeless section Fermi surface without 
first appearing as points. 
The specific heat exponent of $n = 1.4$ for this case is clearly seen in Fig.~\ref{Fig:topology_plot_c}.
  
Schnyder {\it et al.} \cite{Schnyder2011,Schnyder2012} have shown that 
a number of these phases of cubic noncentrosymmetric 
superconductors possess non-trivial topological quantum numbers \cite{topologicalsuperconductors,Beri2010}, 
and have corresponding surface edge states of topological nature. 
In principle, these surface states can be probed experimentally by electron tunnelling through an interface with a normal metal \cite{Schnyder2011tunnelling}. 
However, we propose that a more direct, bulk, measurement of this non-trivial
 gap topology can be obtained by measurements of the low temperature heat
 capacity. If the anomalous power laws we have predicted can be observed, then
 it should be possible to detect the phase boundaries between different types of 
 topological gap structures, and hence determine in a bulk measurement the 
 topological quantum numbers associated with different doping regimes. Phases `1'-`5' in Fig.~\ref{Fig:topology_plot_a_b} correspond to those considered by Schnyder {\it et al.} \cite{Schnyder2011,Schnyder2012}.

The anomalous power-laws due to changes of nodal topology are realised exactly
    at each transition. On the other hand, by being close to, but not exactly
    at one of those transitions one expects a  crossover from conventional
    (fully gapped or normal power-law) behaviour to anomalous behaviour as the
    temperature is increased through a characteristic temperature scale $T^*$. A demonstration of this is provided for the case
    of Li$_{2}$Pd$_{3-x}$Pt$_{x}$B --see Fig.~\ref{Fig:topology_plot_c}, specifically the curves corresponding to $A=3,B=3.5-4.0$ (top panel) and the density plot (bottom panel) showing how the temperature scale $T^*$ converges to zero at the transition. This
    may be used to detect topological transitions in superconductors and in fact shows that the influence of the topological transition extends over a wide region in the phase diagram, not just at a single point.
 The detection of the phase boundary itself is facilitated by its higher specific heat, at fixed low temperature, due to the enhanced phase space available to quasi-particles in that state. This is illustrated in the inset to Fig.~\ref{Fig:topology_plot_c}. We emphasize, however, that it is in the anomalous power-laws that we find the evidence of the special nature of quasi-particles at the phase boundary.

 In recent years it has become widely recognised that point or line nodes in the bulk spectrum of topological matter may evolve into topolgically-protected surface or vortex core states with zero dispersion \cite{Heikkila2011a,Heikkila2011b}. In contrast, the shallow dispersions discussed here exist in the \emph{bulk} excitation spectrum near topological transitions.

\label{sec:conc} To conclude, we have shown that a more general classification
of  nodal characters of superconductors yields anomalous, non-integer power
laws  for low-temperature thermodynamic quantities. Some such power laws can be
used to detect line node crossings; others are expected at  phase transitions
where the nodal topology reconfigures itself. An example is provided by the
topological transitions expected to occur as a function of doping in the
non-centrosymmetric cubic superconductor Li$_{2}$Pd$_{3-x}$Pt$_{x}$B.  It is
important to stress that the observation of anomalous low-temperature power
laws, just as the ordinary, integer ones, requires going to temperatures well
below $T_{c}$. On the other hand, it is not necessary to sit exactly at the transition, as there is a crossover temperature separating conventional from anomalous power-law behaviour that converges to zero there. In that sense the influence of the topological transition extends throughout the bulk phase diagram in a manner similar to that of a quantum critical endpoint \cite{Grigera2001,Vojta2011}.

\begin{acknowledgments}
This works was supported by EPSRC and STFC (U.K.). J.Q. gratefully acknowledges funding from HEFCE and STFC through the South-East Physics network (SEPnet).
\end{acknowledgments}

\begin{appendix}


\section{\label{sec:methods}Calculation of specific heat of nodal superconductors}	

To describe the energy $E_{\bf k}$ of quasi-particles near any specific point on the Fermi surface it is convenient to change coordinate systems to three mutually orthogonal ${\bf k}$-vector components: one perpendicular to the Fermi surface, $k_{\perp}$, and two parallel to it $k_{x\parallel}$, $k_{y\parallel}$. We then have $E_{\bf k}^{2}= {v_F}^2 {k_{\perp}}^2 +  {\Delta_{\bf k}}^{2}$ where $v_F$ is the Fermi velocity, $k_{\perp}=0$ when $\left|{\bf k}\right|=k_F$ (the Fermi vector), and $\Delta_{\bf k}$ is a function determined by the superconducting gap function,  which may be either singlet, triplet or a combination (for non-centrosummetric systems). 

The gap energy spectra shown in Fig.~1 (main text) are described by an equation for the gap energy $\Delta_{\bf k}$ which has the form   $\Delta_{\bf k}^{2}= I_{\mu \nu} k_{\mu\parallel} k_{\nu\parallel} +...$  near to a gap nodal point,  where $I_{\mu \nu}$ is a positive tensor of rank 2 which is determined by the symmetry  of the gap function around that nodal point.  In general $I_{\mu \nu}$  will have two  principal axes, which we can choose as $k_{x\parallel}$ and $k_{y\parallel}$.    If the two eigenvalues corresponding to these axes, $I_x$ and $I_y$ are  both positive and non-zero we have an ordinary point node  with linear spectrum illustrated in Fig.~1 (a) (main text)  (with $I_1=\sqrt{I_xI_y}$). If one eigenvalue is positive and non-zero ($I_1$) and the other is zero we have an ordinary line node.  In the case of two zero    eigenvalues it is necessary  to continue the expansion to quartic or higher order, giving the various shallow nodes or crossings illustrated in Fig.~1 (main text). In principle even higher powers may be possible, but we have not found any realistic examples of gap models where this could occur. In contrast, all the node types presented in Fig.~1 (main text)   can be realised in the specific gap models presented in this paper.

Computation of the density of states integral $g(E)=\int \int \int \delta(E_{\bf k}-E)dk_{x}dk_{y}dk_{z}$ for each of these gap energy spectra gives a unique expression. The specific heat capacity in the superconducting state is in turn given by \cite{Leggett79}
\begin{widetext}
\begin{equation}
C_{v} \equiv  T\left ( \frac{d S}{d T}\right )  
=\sum_{\bf k}\frac{1}{2}k_{B}\beta ^{2}\left[E_{\bf k} + \beta \left ( \frac{d E_{\bf k}}{d \beta} \right )\right] 
E_{\bf k}\sech^{2}\left(\frac{\beta E_{\bf k}}{2}\right)
\label{eq: cv_from_leggett}
\end{equation}
\end{widetext}
where $\beta=1/k_B T$. The low $T$ specific heat capacity can therefore be found analytically from $g(E)$ by evaluating the integral
\(
\frac{1}{2}k_{B}\beta^{2}\int_{0}^{\infty }dE~g(E) 
E^{2} 
\sech^2\left(\beta E/2\right),
\)
giving the power law dependence of a low $T$ specific heat capacity measurement. We find the terms listed in Fig. ~1 (main text) for point and line nodes with linear and non-linear excitation spectra, where $L$ is the length of the line node from end to end. For cases where line nodes cross we expect the terms listed in Fig.~1(main text), where $L$ is now the length of the line node on the Fermi surface measured from the centre to the corner of the crossing whose contribution is being calculated.
The cases in Figs. 1 e) and f) (main text) are not power laws, however they approximate very closely to the non-integer power laws indicated, with very slight variations depending on the constant $L$. In the case of line reconnection transitions such as Fig. 4, plots 3, 8, and 14 (main text), where the Fermi surface includes  regions where the gap energy spectrum expands as  $\Delta_{\bf k}^{2}=I_{1}{k_{x\parallel}}^{2}{k_{y\parallel}}^{2}$, the contribution of the line crossing to the density of states  is $ \sim E^{0.8}$. 


The low-$T$ specific heat capacity can also be calculated numerically using Eq.~(\ref{eq: cv_from_leggett}) by summing over a sufficient number of integration points corresponding to different values of the gap in k-space near the Fermi surface for selected temperature ranges below T$_{c}$. The exponent of temperature, $n$, can then be obtained using the general formula $n=\mathrm{d}\ln C_v/\mathrm{d}\ln T$. When this quantity approaches a constant we have $C_v \propto T^n$. This yields the curves in  Figs.~2 and 4~(c) (main text), where the temperature dependence of the gap is taken to be given by the expression from Ginzburg-Landau theory, $\left(T-T_c\right)^{1/2}$. This approximation, made for convenience, introduces a weak additional temperature-dependence for $T \ll T_c$ compared to the BCS approximation and therefore provides an upper bound on the temperature dependence introduced by the gap. We find that the exponents are not affected by this: even in the case where the temperature dependence of the gap is assumed to be constant, the same power laws are obtained with flat temperature dependence.

Each contour in Fig.~4~(a)(main text) corresponds to a line where the gap energy $\Delta_{\bf k}$ derived from the gap function of Eqs.~(2,3)(main text) vanishes. This gap energy is given by $\left|\Delta_{\bf k}\right| = \left |\Delta_0 \pm d\left({\bf k}\right) \right |=0$, where $d\left({\bf k}\right)=\left|{\bf d}_{\bf k}\right|$. The phase diagram in Fig.~4~(b) (main text) shows the values of the parameters $A$ and $B$ where the nodal topology changes for this form of the gap energy. This form of $\Delta_{\bf k}$ has also been used to compute the specific heat exponent in Fig.~4~(c) (main text). 

\section{Spectrum of $O$-group superconductors} 

Here we give the details of the derivation of the nodal topologies and phase diagram presented in Fig.~4~(b) of the main text. We use a single band mean-field model with antisymmetric spin-orbit coupling to obtain the resulting gap node topologies on the Fermi surface, which we take to be spherical for simplicity. The Bogoliubov-de Gennes Hamiltionian is 

\begin{eqnarray} \label{eq:bdg_ham}
H(\mathbf{k})=\begin{pmatrix}
\hat{h}(\mathbf{k}) &\hat{\Delta} (\mathbf{k}) \\ 
 \hat{\Delta}^{\dagger}  (\mathbf{k}) & -\hat{h}^{T}(-\mathbf{k})
\end{pmatrix}.
\end{eqnarray}
The noncentrosymmetric (NCS) gap function which mixes singlet and triplet as a result of the parity violation in the crystal lattice structure is given in Eq. (1) of the main text,
where $\sigma_{x,y,z}$ denotes the Pauli matrices
and $\Delta_{0,{\bf k}}$ and $\mathbf{d}_{\mathbf{k}}$ denote the singlet and triplet components of the gap respectively.

The normal state Hamiltonian $h(\mathbf{k})$ given in \cite{Schnyder2011} for non-interacting electrons in a crystal without inversion center is
\begin{eqnarray} \label{eq:norm_ham}
h(\mathbf{k})= \varepsilon_{\mathbf{k}}I+\boldsymbol{\gamma}_{\mathbf{k}}\cdot \boldsymbol{\sigma}  
\end{eqnarray}
where $\varepsilon_{\mathbf{k}}=\varepsilon_{\mathbf{-k}}$ is the non-relativsitic metallic dispersion energy. The second term represents the antisymmetric SOC where the coupling constant $\boldsymbol{\gamma}_{\mathbf{k}}=-\boldsymbol{\gamma}_{\mathbf{-k}}$. We will assume the $d$-vector $\mathbf{d}_{\mathbf{k}}$ to be parallel to $\boldsymbol{\gamma}_{\bf k}$. This ensures that SOC is not destructive to the triplet component of the NCS gap function \cite{Frigeri04,Schnyder2011}. 

In a general Ginzburg-Landau Theory the possible symmetries of the superconducting instability depend on the symmetry of the normal state. In this case the symmetry of the normal state is given by $G_{N}= O \times T \times U(1)$ where $O$ denotes the NCS cubic space group, $T$ the time reversal symmetry and $U(1)$ the gauge symmetry operation. In any superconducting transition the $U(1)$ symmetry is broken however other symmetries of the normal state may be broken as well. 
For the purposes of our analysis we will, following \cite{Schnyder2011}, look at the case where only the gauge symmetry and no further symmetries of the crystal lattice are broken i.e.~the superconducting instability has the symmetry of the $A_{1}$ irrep of the $O$ group. This means that the gap function, given in general by equation (1) of the main text, 
takes the specific form of Eqs. (2,3) of the main text. 

With $\mathbf{d}_{\mathbf{k}}$ and $\boldsymbol{\gamma}_{\bf k}$ explicitly defined the Bogoliubov de-Gennes (BdG) Hamiltonian can be diagonalized.
\label{sec:A_1} The single band BdG hamiltonian with SOC defined parallel to the $\mathbf{d}(\bf{k})$ component of the gap \cite{Frigeri04,Schnyder2011} has the property that splitting of the Fermi surface occurs as a result of SOC whereas zero energy excitations occur as a result of the combination of singlet and triplet components of the gap function.  In this case the diagonalized BdG Hamiltonian has eigenvalues of the form 
\begin{eqnarray} \label{eq:eigenvalues}
E=\pm \sqrt{(\varepsilon_{\mathbf{k}} -\mu \pm \left |\boldsymbol{\gamma}_{\bf k}\right |)^{2}+\left |\Delta_{0} \pm |{\bf d}_{\bf k}|  \right |^{2}},
\end{eqnarray}
where the $\pm$ signs inside the square root always take the same value while the sign in the front varie independently. The SOC term $\boldsymbol{\gamma}_{\bf k}$ in this expression has the effect of splitting Fermi sheets whereas the second term $|\Delta_0 \pm |\boldsymbol{d}_{\mathbf{k}}| |$ determines the node topology of the superconducting gap on either of the Rashba-split Fermi sheets.

The zeros of the term $\left|\Delta_0 \pm |d_{\mathbf{k}}|  \right |$ are the locii of zero energy excitations on the Fermi surface. In order to ensure that the SOC parameter does not significantly warp the spherical Fermi surface and therefore change the node topology we assume that $\epsilon_{\bf k} \gg \boldsymbol{\gamma}_{\bf k} \gg \Delta_0$. This hierarchy of energy scales is reasonable to assume as $\epsilon_{\bf k}$ is of the order of eV, $\Delta_0$ is of the order of meV, and observed SOC-induced Fermi surface splittings are small relativistic corrections to $\epsilon_{\bf k}$.

The conditions for zeros in the gap energy spectrum for each Rashba-split Fermi sheet are  $\Delta_0 - |\mathbf{d}_{\bf k}| =0$ and $\Delta_{0} + |\mathbf{d}_{\bf k}|=0$. Thus, one of the Fermi surface sheets is always fully-gapped, while the other one may be nodal. The gap function in this latter case is best normalized with respect to the singlet gap energy:
\begin{eqnarray} 
\Delta_{0}^{2} \left|1 - \frac{|\mathbf{d}_{\bf  k}|}{\Delta_{0}}  \right |^2=\Delta_{0}^{2} \left|1 - |\mathbf{d}'_{\bf k}|  \right |^2=0
\end{eqnarray}
The term $\frac{|\mathbf{d}_{\bf k}|}{\Delta_{0}}$ has here been replaced with the term $|\mathbf{d}'_{\bf k}|$ whose coefficients are scaled by a factor of $\Delta_{0}$. This expression can be fully expanded and used to derive the topological phase boundaries by substituting in the vector $\mathbf{d}'(k)=[AX - BX(Y^{2}+ Z^{2}),AY - BY(Z^{2}+X^{2}),AZ - BZ(X^{2}+Y^{2})]$. A coordinate transformation to spherical polar coordinates using  $X=\cos(\phi)\sin(\theta)$, $Y=\sin(\phi)\sin(\theta)$ and $Z=\cos(\theta)$ allows us to select specific points on the Fermi surface and find solutions in A and B. The phase boundaries of Fig.~4~(b) of the main text separate regions of constant gap topology which were obtained for critical values of the polar and azimuthal angles corresponding to the points where nodes first appear on the gap or where line nodes reconnect at a topological transition.

Phases 1-5 in Fig.~4~(a) of the main text correspond to those considered by Schnyder et al. \cite{Schnyder2011,Schnyder2012}.  The coordinates $\theta=\cos^{-1}\left(1/\sqrt{3}\right) ,\phi= \frac{\pi}{4}$ correspond to one topologically distinct set of point nodes on the spherical Fermi surface shown in Fig.~4~(a), plot 5 of the main article. Substituting these into the equation $\Delta_{0} - |\mathbf{d}_{\bf k}|  =0$ gives the relationship between coefficients `A' and `B': $A = \frac{2}{3}B \pm 1$. The line reconnection shown in Fig.~4~(a), plot 3 of the main article corresponds to the coordinates $\theta=\frac{\pi}{4} , \phi= 0$ and produces solutions $A = \frac{B \pm 2}{2}$. The final phase boundary in Fig.~4~(b) corresponds to point nodes at the coordinates $\theta= \frac{\pi}{2},\phi= 0$ shown in Fig.~4~(a), plot 1, which produces the solutions $A=\pm 1$.

\end{appendix}



\end{document}